\documentclass[showpacs,aps,amsmath,amssymb,twocolumn]{revtex4}
\usepackage{graphicx}
\usepackage{dcolumn}
\usepackage{bm}
\usepackage{psfig}
\usepackage{graphics}
\usepackage{amsmath}
\begin{document}

\title{A liquid crystal analogue of the cosmic string}

\author{Caio S\'atiro}
\affiliation{Departamento de F\'{\i}sica, CCEN,  Universidade Federal 
da Para\'{\i}ba, Cidade Universit\'{a}ria, 58051-970 Jo\~ao Pessoa, PB, Brazil}
\author{Fernando Moraes}  
\affiliation{Laborat\'orio de F\'{\i}sica Te\'orica e Computacional,
Departamento de F\'{\i}sica, Universidade Federal de Pernambuco,
50670-901 Recife, PE, Brazil}
\affiliation{Departamento de F\'{\i}sica, CCEN,  Universidade Federal 
da Para\'{\i}ba, Cidade Universit\'{a}ria, 58051-970 Jo\~ao Pessoa, PB, Brazil}

\begin{abstract}

We consider the propagation of light in a anisotropic medium with a topological line defect in the realm of geometrical optics. It is shown that the effective geometry perceived by light  propagating in such medium is that of a spacial section of the cosmic string spacetime.  

\end{abstract}
\maketitle

Experimentally realizable analogues of cosmic objects have become an important issue in the past few years \cite{analogues,review}. They are mathematically described by some of the same equations and thus make it possible to experimentally check some hypotheses otherwise not accessible to current technology. In particular, the use of an effective geometry for propagating electromagnetic waves has recently been very helpful in devising a analogue black hole \cite{lorenci}. Another analogue system that has been compared with the cosmic string is the irrotational hydrodynamical vortex \cite{visser}. Propagation of phonons in a medium with such vortices can be described by an effective Riemannian geometry which asymptotically agrees with that of the massless spinning cosmic string. The phonon rays are geodesics in this effective geometry and a converging lens behavior has been found \cite{visser}.

\begin{figure}[!h]
\begin{center}
\includegraphics[height=2in]{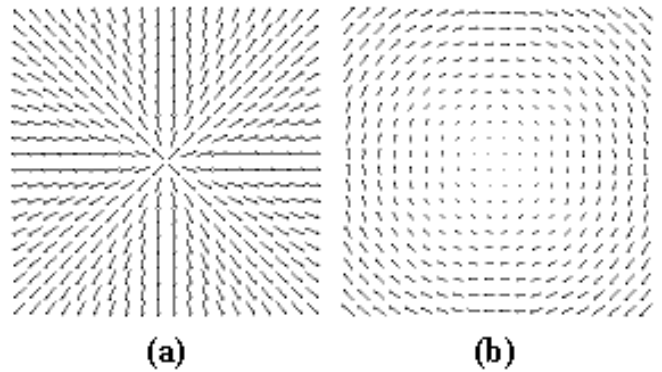} 
\caption{View from above of the string defects in a nematic liquid crystal (a)  disclination,  (b)  vortex.}
\label{campos}
\end{center}
\end{figure}

A nematic liquid crystal is composed of molecules shaped as rods. At high enough temperatures it is in the isotropic phase, a liquid with no orientational or positional order and consequently, global SO(3) symmetry. Lowering the temperature, the overall symmetry is broken and the nematic phase  is reached. This is a mixed phase in the sense that the position of the molecules is still disordered but there is orientational order. The molecules are aligned parallel to each other and the symmetry of the system is therefore SO(2). Stringlike relics of the former, high temperature phase, may appear in the phase transition. These are topological defects like the disclination and the vortex depicted in Fig. (\ref{campos}). Disclinations in nematic liquid crystals have been compared to cosmic strings since it has been shown experimentally \cite{bowick} that their formation obeys the Kibble mechanism \cite{kibble}, originally used to explain the birth of cosmic strings.  

In a 1919 article \cite{grand}, Grandjean demonstrated  that light propagating through a liquid crystalline medium with topological defects like the ones shown in Fig. \ref{campos} is deflected by the defect like it would by a diverging (disclination) or converging (vortex) lens. The light paths are remarkably similar to the geodesics around topological defects in elastic solids that have much in common with cosmic strings \cite{def}. More recently, Joets and Ribotta \cite{joets} proposed a Finsler geometry model to study the propagation of light in anisotropic media like the nematic liquid crystal. This model is the starting point of this work. 

\begin{figure}[!h]
\begin{center}
\includegraphics[height=1.5in]{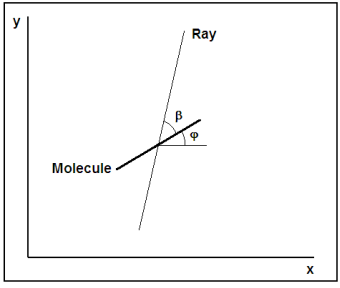} 
\caption{Relative orientation of the nematic molecule and the light ray.}
\label{ray}
\end{center}
\end{figure}

The refractive index for light traveling in a medium with rodlike molecules depends on the angle $\beta$ the ray makes with the molecule as (see Fig. \ref{ray}) 
\begin{equation}
N=\sqrt{n_o^2cos^2 \beta+n_e^2sin^2\beta},\label{N} 
\end{equation}
where $n_o$ and $n_e$ are the so called ordinary and extraordinary refractive index, respectively. $\beta$ is the angle between the optical axis and the ray direction.
The refractive index $N$, therefore, depends on the local orientation of each molecule. 
From Fig. \ref{ray} we have that $\beta$  in terms of the angle $\varphi$, that the molecule makes with the $x$-axis, is
\begin{equation}
\beta(x,y,dx,dy)=\arctan\left(\frac{dy}{dx}\right)-\varphi(x,y),\label{beta} 
\end{equation}
where $\arctan\left(\frac{dy}{dx}\right)$ gives the angle the light ray makes with the $x$-axis at the point of contact with the molecule.

In an inhomogeneous, isotropic, medium Fermat's principle, stating that the optical length of a ray between two given points be the shortest one joining these points,  is nicely interpreted as a geodesic in a Riemannian space \cite{born}. It is then quite natural to model the anisotropic case by Finsler geometry. That is what was done in reference \cite{joets}. In this report we show that, at least for the  two configurations above, the geometry is not only Riemannian but also that of a cosmic string.

Starting with the more general, Finsler geometry case, we identify the element of optical length
\begin{equation} 
ds = N(x^m , dx^{n})\|d{\bf x}\| , \label{ds}
\end{equation} 
with the Finslerian line element
\begin{equation} 
ds = F(x^m,\dot{x}^n)dt, 
\end{equation}
where $F$ is the Finslerian function and $\dot{x}^n$ the derivative of the coordinate $x^n$ with respect to the arc length $t$ of the curve $x^n(t)$, which is the trajectory of the light in the medium. Therefore, the Finslerian function is given by
\begin{equation}
F=N(x^m , dx^{n})\|{\bf \dot{x}}\|. \label{F}
\end{equation}
The metric tensor is obtained from the square of the Finslerian function as \cite{finsler}
\begin{equation}
g_{ij}=\frac{1}{2}\frac{\partial^2 F^2}{\partial  \dot{x}^i \partial  \dot{x}^j}. \label{g}
\end{equation}

The symmetry of the configurations studied here makes it natural to work with cylindrical coordinates. Also, translational invariance along the $z$-axis allows us to make a two-dimensional study. Or, alternatively, we restrict ourselves to the light paths in the plane orthogonal to the defect line.  We define,
\begin{equation}
A=\frac{dy}{dx}=\frac{\dot{y}}{\dot{x}}=\frac{\dot{r}\sin\theta + r\dot{\theta}\cos\theta}{\dot{r}\cos\theta - r\dot{\theta}\sin\theta},\label{A} 
\end{equation}
where the dots represent derivatives with respect to $t$. After some trigonometric manipulation of equations (\ref{beta}) and (\ref{A}), we obtain
\begin{equation}
\sin\beta=\frac{A\cos\varphi-\sin\varphi}{\sqrt{1+A^2}} \label{sin}
\end{equation}
and
\begin{equation}
\cos\beta=\frac{A\sin\varphi+\cos\varphi}{\sqrt{1+A^2}}. \label{cos}
\end{equation}

We find the orientation field $\varphi$, for both cases, by inspection of Fig. \ref{campos}: 
\begin{equation}
\varphi=\theta,\label{fi1}
\end{equation} 
for the disclination, and 
\begin{equation}
\varphi=\frac{\pi}{2}+\theta,\label{fi2}
\end{equation}
for the vortex. 

Now, combining (\ref{sin}) and (\ref{cos}) with (\ref{N}) and (\ref{F}), and with
\begin{equation}
\|{\bf \dot{x}}\|=\sqrt{\dot{r}^{2}+r^{2}\dot{\theta}^{2}},
\end{equation}
and after a lengthly calculation, we find for the square of the Finslerian function the surprisingly simple results
\begin{equation}
F^2=(n_o^{2}\dot{r}^2+n_e^{2}r^{2}\dot{\theta}^2),
\end{equation}
for the disclination, and
\begin{equation}
F^2=(n_o^{2}r^{2}\dot{\theta}^2+n_e^{2}\dot{r}^2),
\end{equation}
for the vortex.

From the above expressions and from (\ref{g}), we get the following metric for the disclination
\begin{equation}
g^{disc}_{ij}=\left(
\begin{array}{cc} 
n_o^2&0\\
0& n_e^2 r^2 
\end{array}\right).
\end{equation}
Rescaling the coordinate $r$ to $R=n_{o}r$ and with $\alpha_{disc}=n_e/n_o$, we get
\begin{equation}
g^{disc}_{ij}=\left(
\begin{array}{cc} 
1&0\\
0& \alpha_{disc}^{2} R^2 
\end{array}\right).
\end{equation}
We have exactly the same result for the vortex, except that  $\alpha_{vortex}=n_o / n_e$. These  conical metrics are obviously Riemannian. In fact, they can be obtained simply by equating the element  of optical length (\ref{ds}) to the Riemannian line element $ds^2=g_{ij}dx^{i}dx^{j}$. This non-Finslerian behavior is  due to the high degree of symmetry of the defects studied here but, it is not necessarily true in the general case, as shown in \cite{joets}.

On the other hand, the cosmic string spacetime is described by the line element \cite{vilenkin}
\begin{equation}
ds^{2}=-c^{2}dt^{2}+dz^{2}+dR^{2}+\alpha^{2}R^{2}d\theta^{2},
\end{equation}
where $\alpha^2=1-8G\mu<1$, with $G$ being the gravitational constant and $\mu$ the linear mass density of the string.  Therefore, light traveling in a nematic with a disclination or a vortex feels an effective Riemannian geometry which  is identical to that of a transversal section of the cosmic string spacetime. The light will travel along geodesics of this effective geometry. Since, in general, for nematics, $n_{o}<n_{e}$, which makes $\alpha_{vortex}<1$  and $\alpha_{disc}>1$, these analogue systems mimic both the ordinary, positive mass density ($\mu>0$) and the more exotic, negative mass density ($\mu<0$), cosmic strings.  Similar defects appear also in the elastic continuum approximation for solids \cite{bjp}. The geodesics in such media have been studied in \cite{disl} and \cite{def}. 

\textbf{Acknowledgments:}
This work was partially supported by PRONEX/FAPESQ-PB, CNPq and CAPES (PROCAD). We are indebted to C. Furtado, T. Paschoal and J. Schaum for many helpful advices.

\end{document}